\documentclass{article}
\title{Hardness of approximating the weight enumerator of a binary linear code}
\author{M.~N.~Vyalyi\thanks{Partially supported by RFBR grant 02-01-00547}\\ 
Institute for Quantum Information\\
California Institute of Technology\\
vyalyi@mccme.ru}
\date{\today}

\usepackage{amsfonts,amssymb,amsmath,amsthm}
\usepackage{fullpage}

\newtheorem{theorem}{Theorem}
\newtheorem{cor}{Corollary}

\let\epsilon\varepsilon
\let\eps\varepsilon
\let\al\alpha

\let\ph\varphi
\let\om\omega

\def\FF{{\mathbb F}_2}

\renewcommand*\P{\ensuremath{\mathrm {P}}}

\newcommand*\PH{\ensuremath{\mathrm {PH}}}
\newcommand*\NP{\ensuremath{\mathrm {NP}}}

\newcommand*{\poly}{\mathop{\mathrm{poly}}}
\newcommand*\GapP{\ensuremath{\mathrm {GapP}}}

\def\ket#1{|#1\rangle }
\def\bra#1{\langle #1|}

\def\CNOT{\mathop{\mathrm{CNOT}}}

\def\monom{\mathop{\mathrm{M}}}

\def\calB{{\cal B}}
\def\tw{{\tilde w}}

\def\WFE{\mathrm{WE}_\om(\al)}
\def\WFEq{\mathrm{WE}(\al)}
\def\wreath{\mathbin{\wr}}
\begin{document}
\maketitle

\begin{abstract}
We consider the problem of evaluation of the weight enumerator
of a binary linear code. We show that the exact evaluation is hard
for polynomial hierarchy. More exactly, if $WE$ is an oracle answering
the solution of the evaluation problem then $\P^{WE}=\P^\GapP$. 
Also we consider  the approximative evaluation of the weight
enumerator. In the case of approximation with additive accuracy
$2^{\alpha n}$, $\alpha$ is constant the problem is hard in the above
sense. We also prove that approximate evaluation 
at a single point $e^{\pi i/4}$ is hard for 
$0<\al<\log_2|1+\om|\approx0.88$.
\end{abstract}

The weight enumerator of a linear binary code $C$ can be defined as follows
\begin{equation}\label{WECdef}
w_C(q)=\sum_{x\in C}q^{|x|}.
\end{equation}
Here $|x|$ is a Hamming weight of binary string $x\in\{0,1\}^n$.
According to the definition, the weight enumerator is a polynomial in
$q$ with integer coefficients. The exact calculation of any coefficient
of the weight enumerator is a $\#P$-problem. Evaluation of the weight
enumerator at any positive integer point is a $\#P$-problem too. 
As for evaluation at arbitrary point the approximate setting seems more
appropriate. Using approximate evaluation with additive
accuracy of $2^{-\poly(n)}$ at different points it is possible to compute
all (integer) coefficients of the weight enumerator.  The vector of
coefficients is expressed as a linear transform of 
the vector of values in  any set of $n+1$ point. A matrix of this
transform is inverse of the Vandermonde matrix. The
accuracy of $2^{-\poly(n)}$ for values of a polynomial degree $n$ 
is sufficient 
to get the coefficients of the polynomial with $1/2$-accuracy.  

To avoid the difficulties that arise when dealing with arbitrary real
(and complex) numbers we  will substitute the problem of the
high-accuracy
evaluation of
the weight enumerator at arbitrary point by the problem
of computation the coefficients of the weight
enumerator.

\textsc{Weight enumerator problem.} Given a binary linear code $C$
find out the coefficients  of the weight enumerator $w_C(\om)$.

We denote an oracle answering the weight enumerator problem by $WE$.

Given coefficients of the weight enumerator of a code it is easy to find
out the minimal distance of the code. The latter problem is
\NP-complete~\cite{Vardy}. Moreover, it is hard to approximate the
minimal distance~\cite{DMS99}. So, the evaluation of the weight
enumerator is \NP-hard.
Here 
we give more exact characterization for the complexity of this problem. 
We show that an oracle answering a solution for the evaluation of the
weight enumerator problem can simulate any oracle in the class
\GapP~\cite{FFK94}. The class \GapP{} is closure of \#P under subtraction.
The formal definition of the \GapP{} uses the notion of counting machine.
\emph{A counting machine} is a non-deterministic Turing machine running in
polynomial time and finishing at either accepting
or rejecting halting state. Given an input word $x$ a counting machine
produces a gap $g_M(x)$. \emph{The gap} is the difference between the number of
accepting computation paths and the number of rejecting computation paths.
The inclusion $\#\mathrm{P}\subseteq\GapP$ and Toda's theorem
$\PH\subseteq\P^{\#\mathrm{P}}$ imply that $\P^\GapP$ is hard for the
polynomial hierarchy.

\begin{theorem}\label{PPtoEXACTWF}
$\P^{WE}=\P^\GapP$.
\end{theorem}

In our reduction we use the techniques of quantum computations. It was
proved by Fenner, Green, Homer and Pruim~\cite{FGHP99} that
determining acceptance possibility for a quantum computation is hard
for the polynomial hierarchy. Here we take the similar approach
considering the problem of very tight approximation of a matrix element
$\bra0U\ket0$ for a quantum circuit realizing an unitary operator~$U$.

It will be shown below that quantum circuits can compute the
approximation of weight enumerator for some codes with an additive accuracy
$2^{\alpha n}$, where $\alpha$ is constant. So, it is
interesting to characterize the complexity of approximate evaluation of
the weight enumerator with an additive  accuracy of $2^{\alpha n}$.
We consider two problems related to this question.

\textsc{Estimation at an arbitrary point.} 
The input of this problem is
$(C,q)$, where $C$ is an $[n,d]$ binary linear code and $q$ is a 
complex number $e^{i\pi\ph}$ with $\ph$ be a rational number.
The output $\tw$ must satisfy the condition
\begin{equation}\label{cond1}
|\tw-w_c(q)|<2^{\al{}n}.
\end{equation}

\textsc{Estimation at the point~$\om$.} 
The input of this problem is
$(C,q)$, where $C$ is an $[n,d]$ binary linear code and $\om=e^{i\pi/4}$.
The output $\tw$ must satisfy the condition
\begin{equation}\label{cond2}
|\tw-w_c(\om)|<2^{\al{}n}.
\end{equation}

We denote the oracles answering these problems by $\WFEq$ and  $\WFE$
respectively.

We will prove that the $WE$ is Turing reducible to
both $\WFEq$ and  $\WFE$. In the latter case we need
an additional restriction
to the constant $\alpha$: 
$0<\al<\al_0=\log_2|1+\om|\approx0.88$.

\begin{cor}\label{PPtoWFEq}
$\P^{\WFEq}=\P^\GapP$
for any $0<\al<1$. 
\end{cor}

\begin{cor}\label{PPtoWFE}
$\P^{\WFE}=\P^\GapP$
for  $0<\al<\al_0$. 
\end{cor}

These corollaries show that the approximation with exponentially small accuracy of the weight function of
general linear code
 by quantum computations is hardly possible.

For basic facts about quantum computation
we refer to the
book of Nielsen and Chuang~\cite{NiCh}. We will need a very little of coding
theory. The basic definitions can be found in the classical
book of MacWillams and Sloane~\cite{McWSl}.
For details of gap-defined classes see the paper~\cite{FFK94}.

\section{Matrix element problem and proof of Theorem~\ref{PPtoEXACTWF}}

We choose  the basis
$B=\{\CNOT, H, T\}$  for quantum circuits, where
\begin{equation}
\CNOT\colon\ket {a,b}\mapsto \ket{a,a\oplus b};\quad
H=\frac1{\sqrt2}\begin{pmatrix}1&1\\ 1&-1\end{pmatrix};\quad
T=\begin{pmatrix}1&0\\ 0&e^{i\pi/4}\end{pmatrix}.
\end{equation}
It is well-known that
this basis provides universal quantum computation.
Solovay--Kitaev theorem guarantees the fast and effective approximation of any
unitary operator by the operators in this basis.

\textsc{Matrix element problem.} Given a description of a quantum circuit
in the basis $B$ and accuracy threshold $\eps$ find out the
approximation $\tilde u$ of a matrix element $\bra0U\ket0$ with accuracy
$\eps$: $|\tilde u-\bra0U\ket0|<\eps$. Here $U$ stands for an operator
realizing by the circuit. 

Let's denote an oracle answering the matrix element problem by $ME$.

\begin{theorem}\label{ME}
$\P^{ME}=\P^{\GapP}$.
\end{theorem}

To prove Theorems~\ref{PPtoEXACTWF} and~\ref{ME}  we construct three Turing
reductions:  a \GapP{} oracle to  
the oracle $ME$ oracle; then the $ME$ oracle to the $WE$ oracle; then
the $WE$ oracle to a \GapP{} oracle.

\subsection{\boldmath Reduction an \GapP-oracle to the $ME$ oracle}

Let $f$ be a \GapP-function.
We assume that $f$ is computed by a counting machine in the normal
form and the machine makes only nondeterministic moves with branching
degree~2.  Lemma~4.3 from~\cite{FFK94} guarantees that we do not loss the
generality: the gaps computing by  machines in normal form
and general ones differ by factor~2.

Reduction of $f$-oracle to the $ME$ oracle reproduce the argument used
in~\cite{Vya}. Let $q$ be the number of nondeterministic moves along a
computation path. The number does not depend on a choice of computation
path due to the normal form condition. So, we may consider  counting
machine described above 
as a deterministic Turing machine $M$ supplied with an additional
input: \emph{a guess string} $r\in\{0,1\}^q$. Bits of the guess string
determine the choices for the nondeterministic moves. This machine
produces the same gap $g_M(x)$ as the initial one.
 
Now we  transform the machine $M$ 
to a Boolean circuit $C_M$ in the basis $\{\oplus,\wedge\}$ computing 1 for pairs (input,guess) finishing at
the rejecting state and computing 0 for 
pairs (input,guess) finishing at
the accepting state. The size of the circuit $C_M$ is quadratic on running
time of the $M$. For fixed input string $x$ the gap produced by $M$ is
the difference between the number of guess strings $u$ that give the circuit
output~0 and the number of guess strings $u$ that give the circuit output~1.

Hereinafter we will simplify notation and will use $+$ and $\cdot$ instead of 
$\oplus$ and $\wedge$ respectively.

Let $s$ be the size of the circuit $C_M$.
By $z^{(j)}_k$, $1\leq k\leq s$, we denote auxiliary variables of the~$C_M$.
We also assume that the 
 circuit output is given by  the value of the variable $z^{(j)}_{s}$. 
Each
assignment in a circuit has the form $z^{(j)}_k:=a*b$ where
$*\in\{+,\cdot\}$ and $a,b$ are either input or auxiliary variables.
This assignment corresponds the equation
$Z^{(j)}_k=z^{(j)}_k+a*b=0$. Note that the values of input variables and
the guess string ($x,u$) 
determine the values of all auxiliary variables. So, for each $x$ fixed the
number of solutions of the system of equations $Z^{(j)}_k=0$, $1\leq
k<s$, $z^{(j)}_{s}=\zeta$ equals the number of rejecting computation paths
of the counting machine~$M$ for $\zeta=1$ and the number of accepting
paths for $\zeta=0$. 

For a polynomial $p$ over the two-element field $\FF$ we denote by 
$\Delta p$ the number
\begin{equation}\label{Delta}
\Delta p=\sum_{x} (-1)^{p(x)}.
\end{equation}

Now let's introduce polynomials over~$\FF$:
\begin{equation}
\begin{split}
&F^\zeta_x(u,z,v)=\sum_{k=1}^{s-1}v_kZ^{(j)}_k+v_0(z^{(j)}_{s}+\zeta),\quad
F_x=(1+w)F^{(0)}_x+w(1+F^{(1)}_x),\\
&\zeta\in\{0,1\},\space v\in\{0,1\}^s,\space w\in\{0,1\} .
\end{split}
\end{equation}

Let us check that 
\begin{equation}\label{Delta=gap}
\Delta F_x(u,z,v,w)=\Delta F^0_x(u,z,v)-\Delta F^1_x(u,z,v)=2^s g_m(x). 
\end{equation}
The polynomials $F^\zeta_x$ are linear in variables $v_j$. So, if values
$x,u,z$ does not satisfy a system  $Z^{(j)}_k=0$, $1\leq
k<s$, $z^{(j)}_{s}=\zeta$ then all possible assignments to variables $v$
contribute the zero term to the sum~\eqref{Delta} for the polynomial
$F^\zeta_x$.  If values 
$x,u,z$ satisfy the system then $F^\zeta_x=0$ for any assignment of $v$. 
Thus any accepting computation path of $M$ contributes $2^s$ to $\Delta
F^0_x(u,z,v)$  and any rejecting path contributes $2^s$ to $\Delta
F^1_x(u,z,v)$.
The first equation in~\eqref{Delta=gap} follows immediately from the
definition~\eqref{Delta}. 

From the construction described above it is clear that function
$x\mapsto F_x$ is computable in polynomial time.

Now we relate the value $\Delta F_x$ to a value $\bra0U(F_x)\ket0$ for
some unitary operator acting in the space $\calB^{\otimes N}$ of $N$
qubits where $N$ is the total number of variables in the polynomial
$F_x$.

The operator $S_J=\Lambda^J(-1)$
(controlled phase shift)
 corresponds to the monomial
$x_J=\prod_{j\in J}x_{j}$ of~$F_x$. Controlled phase shift is defined as
\begin{equation}
\left\{\begin{array}{@{}ll}
\Lambda^J(-1)\ket{x_1\dots x_n}=-\ket{x_1\dots x_n},\qquad
&x_J=1,\\
\Lambda^J(-1)\ket{x_1\dots x_n}=\phantom{-}\ket{x_1\dots x_n},\qquad
&\text{otherwise}.
\end{array}\right.
\end{equation}

Let
\begin{equation}\label{U(F)}
U(F_x)=\prod_{j=1}^{n}H[j] \prod_{J\in\monom(F_x)}S_J\prod_{j=1}^{n}H[j]
\end{equation}
where $\monom(F_x)$ is the set of the monomials of~$F_x$.

We have
\begin{multline}\label{0U0}
\bra0U(F_x)\ket0=\\=\frac{1}{2^N}\sum_{u_1,\dots,u_N}
\bra{u_1,\dots,u_N} \prod_{J\in\monom (f)}S_J\ket{u_1,\dots,u_N}=
\frac{1}{2^N}
\sum_{u_1,\dots,u_N}(-1)^{\sum_{J\in\monom(f)}u_J}
=\\=
\frac{1}{2^N}\sum_{u_1,\dots,u_N}(-1)^{F_x(u)}=2^{-N}\Delta F_x.
\end{multline}

Now, we approximate each factor in~\eqref{U(F)} 
$U(f)$ with accuracy $\delta=2^{-N-2}/M$  by a quantum circuit in the
basis $B$. Here $M=\poly(n)$ denotes the number of factors
in~\eqref{U(F)}. A matrix element  $\bra0\tilde U\ket0$ of
a product $\tilde U$ of these approximations
is very close to $\bra0U(F_x)\ket0$:
\begin{equation}
|\bra0\tilde U-U(F_x)\ket0|<2^{-N-2}.
\end{equation}
So, we may compute the exact value of $\Delta F_x$ if an 
approximate evaluation of $\bra0\tilde U\ket0$ with accuracy
$2^{-N-2}$ is given.

To complete the proof of this reduction we note that
each factor
in~\eqref{U(F)} acts on 4 bits at most.  So, by the Solovay -- Kitaev
theorem~\cite{KShV,NiCh} the required
efficient approximation  $\tilde U$ can be constructed in
$\poly(n)$ time.

\subsection{\boldmath Reduction the $ME$ oracle to the $WE$
oracle}\label{MEtoWE} 

Consider a quantum circuit in the basis $B$ that realize an operator $U$. 
All computational paths of this circuit have the same scale factor $1/2^{N/2}$
where $N$
is the total number
of Hadamard gates in the circuit.
So,
to compute the value
$\bra0U\ket0$ we need to sum up phases along all computational paths 
that start at $\ket0$ and finish at $\ket0$.
Note that an Hadamard gate that applies last
to the particular qubit doesn't affect the number of  
$\ket0$ to $\ket0$ paths. Let 
$d$ be the number of other Hadamard gates. 
Then there are $2^d$ summands in the expression of $\bra0U\ket0$.
Each summand is indexed by a binary vector $u\in\FF^d$ consisted of
qubit values just after application of the corresponding 
Hadamard gate. 

A phase shift along a computational path
consists of two factors. The 
first describes the action of $T$ operators. It has a form
$$
\prod_{k=1}^{t} \om^{\ell_k(u)},\quad \text{where}\ \om=e^{i\pi/4},
$$
and $\ell_k(u)$ is a $\FF$-linear form. 
The second factor describes phase shifts induced by Hadamard gates. 
It has a form
$$
(-1)^{B(u)},\quad \text{where}\ B(u)=\sum_{j=1}^d b_j(u)u_j.
$$
$\FF$-linear form $b_j$ in this expression is the value of a qubit 
just before application of $j$th Hadamard gate.
Using equations $i^{a\oplus b}=i^ai^b(-1)^{ab}$ and $i=\om^2$ 
the second factor can be rewritten in the form similar to the first factor. 

Finally, we get the following expression of 
the matrix element 
$\bra0U\ket0$:
\begin{equation}\label{MEasWF}
\bra0U\ket0 = \frac{1}{2^{N/2}}\sum_{u\in\FF^d}\prod_{k=1}^n \om^{\beta_k(u)}, 
\end{equation}
where $\beta_k$ are $\FF$-linear forms.

It is clear that 
up to the factor $1/2^{N/2}$
this expression 
equals $w_C(\om)$---the 
value of weight enumerator at the point $\om$ for some
binary linear code~$C$. 
Coefficients of linear forms $\beta_k(u)$ are 
columns of the generator matrix for 
this code $C$.

So, we reduce the matrix element problem to the weight enumerator
evaluation problem. 

\subsection{\boldmath Reduction the $WE$ oracle to a \GapP{} oracle}

Let $C$ be a linear binary code. It is a $d$-dimensional subspace of an
$n$-dimensional space $\FF^n$. 
By usual properties of \GapP{} functions~\cite{FFK94} the function 
$
E(D(C),q)=w_C(q)\in\GapP
$, where $D(C)$ is a description of the code $C$ and $q$ is integer. 
The required reduction follows from the fact that all coefficients of
$w_C(q)$ can be restored efficiently from the value $E(D(C),2^n)$.

\section{An approximate evaluation}

Theorem~\ref{PPtoEXACTWF} establishes the hardness of high-accuracy 
approximation of the weight enumerator. 

Now
we will prove two results concerning the approximation with  moderate accuracy.
Our choice of accuracy bounds is motivated by observation of
Knill and Laflamme~\cite{KL98}  that the power of quantum computation 
is just the ability to produce a Bernoulli sequences with a parameter
$p$ given by a real or imaginary part of a matrix element $\bra0U\ket0$ 
of an operator realized by a quantum circuit.
So,  value of the matrix element can be estimated with an accuracy $n^{-O(1)}$
and exponentially small  error probability. 

The formula~\eqref{MEasWF} shows that the value of weight enumerator $w_C(\om)$
can be evaluated in some cases with rather high accuracy $n^{-O(1)}2^{N/2}$.
Interesting  quantum computation that does not reduced directly to the 
probabilistic one corresponds the case $N<2d/\log d$.

Corollaries~1 and 2 show that the general case differs from the case of
these special codes. In this section we give the proofs of these corollaries.

\subsection{Combining codes: direct sums and wreath sums}

In this section we introduce two operations with linear codes
and describe corresponding
transformations of
weight enumerators.

Consider an $[n_1,d_1]$ linear code $A$ 
 and an
$[n_1,d_1]$ linear code $B$. Let $G_A$ and $G_B$ are the generator
matrices of codes  $A$ and $B$ respectively.

A direct sum $A\oplus B$ is a $[n_1+n_2,d_1+d_2]$ code defined by
generator matrix as follows:
\begin{equation}
\begin{pmatrix}
G_A&0\\ 0& G_B
\end{pmatrix}.
\end{equation} 
The weight enumerator of the direct sum is just the product of weight
enumerators of summands:
\begin{equation}
w_{A\oplus B}(q)=w_A(q)w_B(q).
\end{equation}

A wreath sum
$A\wreath B$  is a $[n_1n_2,d_1+d_2]$ code.
To define 
elements of the
generator matrix $G_{A\wreath B}$  for $A\wreath B$ 
we assume that rows of $G_{A\wreath B}$ are indexed by pairs $(k_1,0)$, 
$1\leq k_1\leq d_1$,
and $(0,k_2)$, 
$1\leq k_2\leq d_2$, and columns are indexed by pairs
$(l_1,l_2)$, $1\leq l_1\leq n_1$, $1\leq l_2\leq n_2$. 
In this setting the elements of $G_{A\wreath B}$
are given by 
\begin{align}
&(G_{A\wreath B})_{(k,0)(l_1,l_2)}=(G_A)_{k_1l_1}\\
&(G_{A\wreath B})_{(0,k)(l_1,l_2)}=(G_B)_{k_2l_2}.
\end{align}
Each pair of words $(a,b)$, $a\in A, b\in B$ determines in natural way
a word $c_{ab}\in A\wreath B$. Each bit $\alpha$ in the codeword $a$ is
extended 
to a string $\alpha^{l_2}$. This string is added by the codeword $b$ to
produce the $l_2$-block of the codeword $c_{a,b}$.
So, for weights of these codewords the following
relation holds
$$
|c_{ab}|=|a|(n_2-|b|)+|b|(n_1-|a|).
$$
From this equation we get the following relation between weight enumerators:
\begin{equation}\label{wreathwf}
w_{A\wreath B}(q)=\sum_{a\in A} q^{n_2|a|} w_B(1+q^{n_1-2|a|}).
\end{equation}

\subsection{Proof of Corollary~\ref{PPtoWFEq}}

We will construct a Turing reduction of 
the weight enumerator evaluation problem
to the $\WFEq$ problem. 

Namely, we will evaluate weight enumerators of 
a code $C_k=C\oplus I_k$ at several point $q_j$. Here $I_k$ means the
trivial code given by identity generator matrix. The parameter $k$ will
be chosen later.

Choose $r$ such that $\beta=2^\al/(2-r)<1$.
Note that $(2^{\al (d+k)})$-evaluation of 
$w_{C_k}$ implies much stronger estimation of $w_C$:
if $|q-1|<r$ and
$
|w_{C_k}(q)-\tw(q)|<2^{\al (d+k)}
$
then
\begin{equation}\left|\frac{w_{C}(q)(1+q)^k-\tw}{(1+q)^k}\right|=
\left|w_{C}(q)-\frac{\tw}{(1+q)^k}\right|<2^{\al d}\beta^k=2^{-\Omega(k)}.
\end{equation}
(In asymptotic notations we assume that $k\to\infty$ while other
parameters are fixed.) 

To compute the coefficients of weight enumerator $w_C(q)$ we
use this tight bound for $n+1$
points inside the circle $|z-1|<r$. The distances between these points
can be chosen greater than $r/(2(n+1))$.
So, the inverse of Vandermonde determinant for these points is  
$2^{O(n^2\log n)}$.

Thus, for some $k=\poly(n,d)$ we can compute the coefficients of the weight
enumerator with accuracy $1/2$.

\subsection{Proof of Corollary~\ref{PPtoWFE}}

Note that reduction in the subsection~\ref{MEtoWE} uses the evaluation
of the weight enumerator at the point $\om$ only.

To construct a Turing reduction of 
the weight enumerator evaluation problem
to $\WFE$ problem
we will obtain an evaluation of the  weight enumerator of a code $C$ 
at point $\om$ using the ($2^{\al (n+a+k)}$)-estimations of  weight enumerators
for codes $C(a,k)=C_a\wreath I_k$ at the same point. Here $I_k$ stands for a
trivial
code of dimension~$k$ and code $C_a$ is produced from the code $C$
by padding $a$ zeroes to each codeword. 
From~\eqref{wreathwf} we get for these codes:
\begin{equation}\label{CwI}
w_{C(a,k)}=\sum_{w\in C}\om^{k|w|}(1+\om^{n+a-2|w|})^k.
\end{equation}
We assume that $k\equiv 1\pmod8$.

Divide the weight enumerator sum for code $C$ in four parts:
\begin{equation}
w_C(\om)=
M_0+M_1+M_2+M_3;\quad
M_j=\sum_{w\in C, |w|\equiv j\pmod4}\om^j.
\end{equation} 
It is clear that $|M_j|<2^n$.

Now rewrite~\eqref{CwI} as
\begin{equation}\label{wM}
w_{C(a,k)}=M_0(1+\om^{n+a})^k+M_1(1-i\om^{n+a})^k+M_2(1-\om^{n+a})^k+
M_3(1+i\om^{n+a})^k
\end{equation}

Take $a$ such that $n+a=1\pmod 8$. Let $\tw_1$ be an 
($2^{\al (n+a+k)}$)-evaluation of the $w_{C(a,k)}$ and $N=|1+\om|$.
From~\eqref{wM} we have
\begin{equation}
|M_0+M_1\om^{-1}+
M_2\frac{(1-\om)^k}{N^k}+
M_3\frac{(1-\om^{-1})^k}{N^k}-\frac{\tw_1}{N^k}|<
2^{\al(n+a+k)}/N^k=2^{\al (n+a)}\left(\frac{2^\al}{N}\right)^k.
\end{equation} 
The last factor is $2^{-\Omega(k)}$. 
Note that 
$$
\frac{(1-\om)^k}{N^k}=2^{-\Omega(k)}
\quad\text{and}\quad
\frac{(1-\om^{-1})^k}{N^k}=2^{-\Omega(k)}.
$$
So, for some $k=\poly(n,-\log\eps)$ we obtain an $\eps$-evaluation
$\mu_0$
 of
$M_0+M_1\om^{-1}$. 

Incrementing $a$ by 2 in the above argument we get $\eps$-evaluations
$\mu_j$ of
$M_j+M_{j+1}\om^{-1}$.
Solving the linear system  
\begin{equation}
\begin{split}
&x_0+x_1\om^{-1}=\mu_0,\\
&x_1+x_2\om^{-1}=\mu_1,\\
&x_2+x_3\om^{-1}=\mu_2,\\
&x_3+x_0\om^{-1}=\mu_3,\\
\end{split}
\end{equation}
we get the $O(\eps)$-approximations for $M_j$. It is enough to solve
the weight enumerator evaluation problem for code $C$.

\end{document}